\documentclass[11pt]{article}
\def\beqar {\begin{eqnarray}}
\def\eeqar {\end{eqnarray}}
\def\beq {\begin{equation}}
\def\eeq {\end{equation}}

\def\half {{\textstyle{1\over 2}}}
\def\Tr {{\rm Tr}}

\def\del {{\partial}}

\def\l{{\lambda}}

\def\bz {\bar{z}}

\def \A {{\cal A}}
\def \D {{\cal D}}

\def \L {{\cal L}}
\def \O {{\cal O}}

\def\no2 {{\textstyle{n\over 2}}}

\begin{document}

\begin{titlepage}
\null\vspace{-62pt}

\pagestyle{empty}
\begin{center}
\rightline{}
\rightline{}

\vspace{1.0truein} 
{\Large\bf Electromagnetic interactions of higher dimensional }\\
\vskip .2in
{\Large\bf quantum Hall droplets}\\
\vskip .3in
\vspace{.5in}DIMITRA KARABALI \footnote{E-mail address: dimitra.karabali@lehman.cuny.edu} \\
\vspace{.2in}{\it  Department of Physics and Astronomy\\
Lehman College of the CUNY\\
Bronx, NY 10468}\\
\end{center}
\vspace{0.5in}

\centerline{\bf Abstract}

Using a $W_{N}$-gauge theory to describe electromagnetic interactions of spinless fermions in the lowest Landau level, where the $W_{N}$ transformations are nonlinear realizations of $U(1)$ gauge transformations, we construct the effective action describing electromagnetic interactions of a higher dimensional quantum Hall droplet. We also discuss how this is related to the Abelian Seiberg-Witten map. Explicit calculations are presented for the quantum Hall effect on ${\bf CP}^k$ with $U(1)$ background magnetic field. The bulk action is a K\"ahler-Chern-Simons term whose anomaly is cancelled by a boundary contribution so that gauge invariance is explicitly satisfied. 
\vskip .1in
\vskip .5in

\baselineskip=18pt

\end{titlepage}

\hoffset=0in
\newpage
\pagestyle{plain}
\setcounter{page}{2}
\newpage

\section{Introduction}

The study of quantum Hall effect in higher dimensions has recently attracted a lot of attention \cite{HZ}-\cite{KN4} following the original work of Zhang and Hu
\cite{HZ}. Similarly to the classic two-dimensional QHE it provides a framework for new ideas about bosonization, topological field theories, noncommutative geometry and D-branes, in addition to the possibility of higher spin theories and its connection to the observed spin-Hall effect \cite{spinhall}.
Zhang and Hu considered the Landau problem for charged 
fermions on $S^4$ with a background magnetic field corresponding to 
the standard $SU(2)$ instanton.
This leads to a many-body picture in terms of incompressible quantum Hall droplets, which give rise to gapless edge excitations, in a way similar to the two-dimensional QHE.

Recently we generalized the Zhang, Hu construction to arbitrary even dimensions by formulating the quantum Hall effect on the even dimensional, complex projective spaces ${\bf CP}^k$ \cite{KN1}. Within this framework, we also presented a method for obtaining the effective action for edge excitations of a quantum Hall droplet in any dimension \cite{KN2, KN3}, based on a collective formulation of the quantum mechanical density matrix in the limit of large number of fermions or equivalently a large droplet \cite{sakita1}.

In this paper we will analyze the electromagnetic interactions (in addition to the uniform magnetic field necessary for the formation of the QHE) of the underlying charged fermions, by coupling the higher dimensional quantum Hall system, restricted to the lowest Landau level, to a weak electromagnetic field. We are particularly interested in deriving the bulk effective action in terms of the external electromagnetic field and its relation to the edge boundary action. 

It is well known that in the case of two dimensions, one can directly integrate out the fermions and derive the bulk effective action which is essentially an Abelian Chern-Simons action with its coefficient given by the quantized Hall conductance. Further this action and the boundary action are related by the property of anomaly cancellation. The Chern-Simons action defined on a space with boundary is not gauge invariant, the non-invariance given in terms of a surface term. The requirement of gauge invariance implies the presence of a boundary action with the appropriate gauge transformation. It was first shown by Wen based on the anomaly cancellation property that the edge dynamics of a two-dimensional Hall droplet is described in terms of massless chiral fields \cite{wen}, \cite{IKS}.

Our approach here is somewhat different. Starting from a very generic matrix formulation of the lowest Landau level Hall droplet dynamics, we explicitly derive the bulk effective action involving the perturbative electromagnetic fields. The electromagnetic interaction to the edge degrees of freedom is also obtained. One can show that the total action, bulk and edge, is explicitly gauge invariant. The strategy we follow is an adaptation of a method used by Sakita \cite{sakita2}, \cite{shizuya} to derive the electromagnetic interactions of LLL electrons in the two-dimensional plane based on the $W_{\infty}$ - gauge field theory \cite{IKS}, \cite{wadia}.

This paper is organized as follows. In section 2 we present a matrix formulation of the dynamics of the lowest Landau level with and without electromagnetic interactions and we outline a general method for deriving the corresponding effective action of a quantum Hall droplet in arbitrary number of dimensions. In the rest of the paper we present an explicit analysis for the case of a quantum Hall droplet on ${\bf CP}^k$ with a $U(1)$ background field. In section 3 we briefly review the structure of the lowest Landau level and the emerging star product for ${\bf CP}^k$ with $U(1)$ background magnetic field. In section 4 we present the derivation of the effective action in the absence of electromagnetic interactions and in section 5 we extend this to include electromagnetic interactions. The role of the Seiberg-Witten map in the derivation of the effective action is discussed. In section 6 we analyze the gauge invariance of the effective action and explicitly show how the anomaly of the bulk part, which turns out to be a K\"ahler-Chern-Simons term, is cancelled by the boundary contribution. A brief conclusion and comments are presented in section 7. 

\section{General approach}

We shall first review the method used in \cite{KN2, KN3} to derive the action describing the dynamics of a generic lowest Landau level Hall droplet in the absence of external weak electromagnetic interactions and then discuss how this gets modified in order to describe electromagnetic interactions and derive the corresponding effective action. 

Let $N$ denote the dimension of the one-particle Hilbert space corresponding to the states of the lowest Landau level, $K$ of which are occupied by fermions. Spin degrees of freedom are neglected, so each state can be occupied by a single fermion. In the presence of a confining potential $\hat{V}$, the degeneracy of the LLL states is lifted and the fermions are localized around the minimum of the potential forming a droplet. The choice of the droplet we are considering is specified by a diagonal density matrix 
 $\hat {\rho}_0$ which is equal to 1 for occupied states and zero for unoccupied states. The most general fluctuations which preserve the LLL condition and the number of occupied states are unitary transformations of $\hat {\rho}_0$, namely $\hat {\rho}_0 \rightarrow \hat{\rho}=\hat{U}  \hat {\rho}_0 \hat{U} ^ \dagger$, where  $\hat{U}$ is an $(N \times N)$ unitary matrix. The
action which determines ${\hat U}$ is given by 
\beq
S_{0}= \int dt~ \left[ i \Tr ({\hat \rho}_0 { \hat U}^\dagger \del_t {\hat U})
~-~ \Tr ({\hat \rho}_0 {\hat U}^\dagger {\hat{V}} {\hat U}) \right]
\label{1}
\eeq
where $\hat {V}$ is the confining potential. We have used the fact that on the LLL the Hamiltonian is $\hat{V}$ up to an additive constant. $\hat{U}$ can be thought of as a collective variable describing all the possible droplet excitations within the LLL. Variation of $\hat{U}$ in (\ref{1}) leads to the extremization condition for $S_{0}$ as
\beq
i {{\partial \hat{\rho}} \over {\partial t}} = [ \hat{V} , \hat{\rho}]
\label{3}
\eeq
which is the expected evolution equation for the density matrix $\hat{\rho}$.  

The action $S_{0}$ can also be written as
\beq
S_{0}= N \int d\mu dt~ \left[ i ({\rho}_0 *{  U}^\dagger  * \del_t { U})
~-~ ({ \rho}_0 *{U}^\dagger  * {{V}} * {U}) \right]
\label{4}
\eeq
where $d\mu$ is the volume measure of the space where QHE has been defined and $\rho_0,~U,~V$ are the symbols of the corresponding matrices on this space. In our notation the hatted expressions correspond to matrices and unhatted ones to the corresponding symbols, which are classical functions on the space where QHE is defined.  

It was shown in \cite{KN2}, that using the star product appropriate for ${\bf CP}^k$ with $U(1)$ background field and the fact that $\hat{U} = \exp{i \hat{\Phi}}$, the action $S_0$ in (\ref{4}) describes a higher dimensional chiral action in terms of the Abelian bosonic field $\Phi$ defined on the boundary of the droplet. 

In order to describe electromagnetic interactions we modify the matrix version (\ref{1}) of $S_0$ by introducing a gauge potential $\hat{\A}$ whose variation is such that the new action is invariant under time dependent $U(N)$ rotations. The new gauged action is
\beq
S= \int dt~ \Tr \left[ i  \hat {\rho}_0  \hat {U}^\dagger \del_t \hat {U}
~-~  \hat {\rho}_0 \hat {U}^\dagger \hat{V} \hat{U} - \hat {\rho}_0 \hat {U}^\dagger \hat{\A} \hat{ U} \right]
\label{6}
\eeq
Under 
\beq
\hat{U} \rightarrow \hat{h} \hat{U} ~,~~~~\hat{\A} \rightarrow \hat{\A}'
\label{7}
\eeq
the action (\ref{7}) changes as
\beq
S \rightarrow S +  \int dt~   \Tr ~ \hat{U} {\hat \rho}_0 { \hat U}^\dagger \left[ i \hat{h} ^\dagger \del_t \hat{h} - \hat{h} ^\dagger \hat{V} \hat{h} + \hat{V} - \hat{h} ^\dagger \hat{\A}' \hat{h} + \hat{\A} \right]
\label{8}
\eeq
For infinitesimal transformations $\hat{U} \rightarrow \hat{U} + \delta \hat{U}~,~ \hat{\A} \rightarrow \hat{\A} + \delta \hat{\A}$,
the action remains invariant if 
\beqar
\delta \hat{\A} & = &  \del _t {\hat{ \l }}-i [\hat{\l}, \hat{V}+\hat{ \A} ] \nonumber \\
\delta\hat{U} & =& - i \hat{\l} \hat{U}
\label{9}
\eeqar
where $\hat{h} = \exp (-i \hat{\lambda})$.
 As before the action $S$ in (\ref{8}) can be written in terms of the corresponding symbols as
 \beq
S= N \int dt~d\mu  \left[ i  \rho_0 * U^\dagger  * \del_t U
~-~ \rho_0 *U^\dagger * V *U- \rho_0 *U^\dagger * \A *U \right]
\label{10}
\eeq
where the ($N \times N$) matrices have been replaced by their symbols, matrix multiplication by the star product,  and $\Tr$ by $N \int d\mu$. The action (\ref{10}) is now invariant under the infinitesimal transformations
\beqar
\delta \A (\vec{x}, t)  & = & \del _t \l (\vec{x}, t) - i \big( \l* (V+\A) - (V + \A) * \l)\big) \label{11a} \\
\delta U & =& - i \l * U
\label{11}
\eeqar
Equations (\ref{11a}) and (\ref{11}) express what we refer to as $W_{N}$ gauge transformations.
 In the case of QHE on the two-dimensional plane, the transformation (\ref{11a}) was identified as the $W_\infty$ gauge transformation \cite{sakita2, wadia, IKS}. If the action $S$ in (\ref{10}) describes the interaction of the quantum Hall system to a weak external electromagnetic field $A_{\mu}$, it has to be invariant under the usual gauge 
 transformation
 \beq
 \delta A_{\mu} = \del_{\mu} \Lambda
 \label{12}
 \eeq
where $\Lambda$ is the infinitesimal gauge parameter. This implies then that the transformation (\ref{11a}) can be thought of as a nonlinear realization of the gauge transformation (\ref{12}). This requirement determines (up to some ambiguities to be discussed later) $\A$ as a function of $A_{\mu}$ and therefore the electromagnetic interaction of the Hall droplet in (\ref{10}). We assume that the energy scale of $A_\mu$ is much smaller than that of the confining potential $V$ and the existing uniform magnetic field $\bar{B}$ so that the restriction to the lowest Landau level is still valid. In this sense we refer to the additional external electromagnetic field as weak. 

The method outlined in this section is quite general and can be applied to a quantum Hall system defined on any space. In the following sections we shall apply this method to the quantum Hall effect defined on ${\bf CP}^k$ with a uniform $U(1)$ background field.

\section{ Specifics of QHE on ${\bf CP}^k$ with $U(1)$ background field}

Here we shall briefly review the structure of the lowest Landau level and the emerging star product for ${\bf CP}^k$, which are the crucial ingredients in constructing the effective action (\ref{10}). We shall mainly follow the presentation in \cite{KN2, KN3} and in the review article \cite{KN4}. 

${\bf CP}^k$ is a $2k$ dimensional manifold parametrized by $k+1$ complex coordinates $v_a$, such that
\beq
\bar{v}_a v_a =1
\label{13}
\eeq
with the identification $v_a \sim e^{i \theta} v_a$. One can further introduce local complex coordinates $z_{\alpha}$, $\alpha=1,\cdots, k$, by writing
\beqar
v_{\alpha} & = & {z_{\alpha} \over {\sqrt{1 + \bz \cdot z}}}~,~~~~~~~\alpha =1, \cdots , k \nonumber \\
v_{k+1} & = & {1 \over {\sqrt{1 + \bz \cdot z}}} 
\label{complex}
\eeqar

The $U(1)$ background magnetic field (which leads to the Landau states) is introduced via a gauge potential
\beq
\bar{A} = -i n  \bar{v} \cdot dv
\label{17}
\eeq
where $n$ is an integer and $n=2 \bar{B} R^2$, where $\bar{B}$ is the magnetic field and $R$ is the radius of ${\bf CP}^k$.  

The volume element on ${\bf CP}^k$ is
\beq
d\mu = {k! \over \pi^k} {{d^{2k} x} \over {(1+ \bz \cdot z)^{k+1}}}
\label{measure}
\eeq
The normalization is such that $\int d\mu =1$.
\vskip .2in

\noindent
$\underline{\rm{Lowest~Landau~Level}}$

The lowest Landau level wavefunctions were derived in \cite{KN2}. They can be written in terms of the local coordinates $z_i$ as
\beqar
\Psi_m(\vec{x}) &=& \left[ {n! \over i_1! i_2! ...i_k!
(n-s)!}\right]^\half ~ {z_1^{i_1} z_2^{i_2}\cdots z_k^{i_k}\over
(1+\bz \cdot z )^{n \over 2}} \nonumber\\
s &=& i_1 +i_2 + \cdots +i_k ,~~~0\le i_i \le n~,   0 \le s \le n \label{wav}
\eeqar
These wavefunctions form a symmetric rank $n$ representation $J$ of $SU(k+1)$. The index $m=1,\cdots,N$ in (\ref{wav}) labels the states within this representation $J$. The dimension of the representation is 
\beq
dim J = {{(n+k)!} \over {n! k!}} \equiv N
\label{dim}
\eeq
Apart from the factor $(1+\bz \cdot z )^{-n/2}$ the wavefunctions (\ref{wav}) are holomorphic. They are the coherent states for ${\bf CP}^k$.
\vskip .2in

\noindent
$\underline{\rm{Star~product}}$

The symbol corresponding to a ($N \times N$) matrix $\hat{O}$, with matrix elements $O_{ml}$, acting on the Hilbert space of the lowest Landau level is a function on ${\bf CP}^k$ defined by
\beq
O(\vec{x}, t) = \sum_{m,l} \Psi_m(\vec{x}) O_{ml}(t) \Psi^*_l(\vec{x})
\label{symb}
\eeq
We found in \cite{KN2}  that for large $n$ the symbol corresponding to the product of two operators is
\beqar
\big(\hat{O}_1\hat{O}_2\big)_{symbol} & \equiv &  O_1(\vec{x}, t) * O_2 (\vec{x}, t) \nonumber \\
& = & O_1 O_2 ~+ ~{i \over n} (\Omega ^{-1})^{\bar{\alpha}\beta}~ {{\del  O_1(\vec{x},t)} \over {\del \bz^{\alpha}}}~ {{\del O_2(\vec{x},t)} \over {\del z^{\beta}}}  + ~{\cal{O}}({1 \over n^2})
\label{star}
\eeqar
where $\Omega$ defines the Poisson bracket on ${\bf CP}^k$, namely
\beq
\{ O_1(\vec{x}, t ), O_2(\vec{x}, t) \} =  (\Omega ^{-1})^{\bar{\alpha}\beta}~ \Big({{\del  O_1(\vec{x},t)} \over \del \bz^\alpha} ~ {{\del O_2(\vec{x},t) } \over \del z^\beta} ~-~ {{\del O_2(\vec{x},t)} \over \del \bz^\alpha}  ~{{\del O_1(\vec{x},t)} \over \del z^\beta} \Big)
\label{poisson}
\eeq
In terms of the complex coordinates $z,~\bz$ we have
\beq
(\Omega ^{-1})^{\bar{\alpha}\beta} = -i (1 + \bz \cdot z ) (\delta ^{\alpha\beta} + \bz ^\alpha z^\beta)
\label{omega}
\eeq
The important result here is that, for large $n$, the symbol of the commutator of two operators acting on the Hilbert space of the lowest Landau level is essentially the Poisson bracket of the corresponding symbols, namely
\beqar
\big([\hat{O}_1,~\hat{O}_2]\big)_{symbol} & = & {i \over n} \{O_1(\vec{x}, t) ,~O_2(\vec{x}, t) \} \nonumber \\
& = & {i \over n} (\Omega^{-1})^{ij} ~{{\del O_1(\vec{x}, t)} \over {\del x^i}} ~ {{\del O_2 (\vec{x}, t)} \over {\del x^j}}
\label{com}
\eeqar
where $x_i$ are the real coordinates of $\bf{CP}^k$ and the indices $i,j$ take values $i,j = 1,\cdots , 2k$.

\section {Edge effective action without electromagnetic interactions}

In this section we review the derivation of the edge effective action in the absence of electromagnetic interactions (besides the uniform magnetic field $\bar{B}$), as given in \cite{KN2, KN3}.

Starting from (\ref{1}) and using $\hat{U} = \exp (i \hat{\Phi})$ we find that, up to a constant, 
\beqar
S_0 = \int dt &&\Tr  \Big[  \Big( - \hat {\rho}_0 
-{i \over 2}  [{ \hat \Phi} ,{\hat \rho}_0 ]
+{1 \over 3!} [{\hat \Phi} ,[{\hat \Phi} ,{\hat \rho}_0]]
+\cdots \Big)~\del_t {\hat \Phi} \nonumber\\
& - &  \Big(  i [{\hat \rho}_0 , {\hat V}] {\hat \Phi} )
+{1\over 2} [{\hat \rho}_0 , {\hat \Phi} ] [ {\hat V}, {\hat \Phi} ])+
\cdots \Big)~\Big] 
\label{s0}
\eeqar
where the first line gives the expansion of the kinetic energy in terms of $\hat{\Phi}$ commutators and the second line gives the potential energy.

We choose $\hat{\rho} _0$ to be the ground state density matrix, so that $[\hat{\rho}_0,~\hat{V}] = 0$. We shall further use a convenient spherical symmetric confining potential $V= V(r^2)$, where $r^2 = \bz \cdot z$, so that $\rho_0$ corresponds to a spherical droplet, although the results can be easily extended to nonspherical droplets.  In the limit $n \rightarrow \infty$, or equivalently $N \rightarrow \infty$, when a large number $K$ of LLL states are occupied, the classical function $\rho_0 (r^2)$ is essentially a step function identifying the droplet. An explicit calculation of $\rho_0$ in the case of ${\bf CP}^k$ with $U(1)$ background field was done in \cite{KN2}. As we mentioned in the previous section, in this case, the LLL states form a symmetric rank $n$ representation of $SU(k+1)$. If we choose a confining potential $\hat{V}$ with an $SU(k)$ symmetry, the ground state is formed by filling up a certain number of complete $SU(k)$ representations, starting with the singlet, the fundamental, rank two and so on, up to, let us say rank $M$ symmetric representation. We found in \cite{KN2} that in the large $N$, large $K$ limit, where $N \gg K$
\beq
\rho_0  =  \Theta \Big(1 - {{n r^2} \over M}\Big) = \Theta \Big(1 - {{R^2 r^2} \over {R_D^2}}\Big)
\label{step}
\eeq
where $\Theta$ is the step function, and $R_D$ is the radius of the droplet, $R_D^2 = {M \over {2\bar{B}}}$ and $n=2\bar{B}R^2$.

Writing $S_0$ as an integral over ${\bf CP}^k$ and using (\ref{com}), we find that in the large $n$-limit, (\ref{4}) becomes
\beq
S_0 = {N \over{2n}} \int dt d\mu \left( \{\Phi, \rho_0\} \del_t \Phi + {1 \over n} \{\rho_0, \Phi\}\{V, \Phi\}\right)
\label{s0cp}
\eeq
For a spherically symmetric $\rho_0$
\beqar
\{ \Phi , \rho_0 \}&=& (\Omega^{-1}) ^{ij} {\del \Phi \over
\del x^i} {\del \rho_0 \over \del x^j}\nonumber\\
&=& (\Omega^{-1})^{ij} {\hat e}_j {\del \Phi \over \del x^i}
{\del \rho_0 \over \del r}\nonumber\\
&=& (\L\Phi ) \left( {\del \rho_0 \over \del r^2}\right)
\label{lder}
\eeqar
where
\beq
\L\Phi = \left[ 2r {\hat e}_j (\Omega^{-1})^{ij}
{\del \Phi \over \del x^i}\right]
\label{8a}
\eeq 
The coordinates $x_i$ are dimensionless, measured in units of the radius $R$ of the compact
space. 
${\hat e}_j$ is a unit vector normal to the boundary $\del\D$ of the
droplet and $r=\sqrt{\sum x_i^2}$ is the dimensionless, normal coordinate. $\L\Phi$
involves only derivatives of $\Phi$ with respect to a tangential direction, which is the Poisson conjugate to the normal to the droplet. Similarly
\beq
\{\rho_0, \Phi\}\{V,\Phi\} = (\L \Phi)^2 {{\del \rho_0} \over {\del r^2}} {{\del V} \over {\del r^2}}
\label{rhoV}
\eeq
Using (\ref{lder}), (\ref{rhoV}) in (\ref{s0cp}) we find 
\beq
S_0 = {N \over 2n}\int dtd\mu \left( {\del \rho_0 \over \del r^2}\right)~
\left[ {\del \Phi \over \del t} (\L\Phi ) 
+{1\over n} {\del V\over \del r^2} (\L\Phi )^2
\right] +{\cal O}(1/n)
\label{12a}
\eeq
where the operator $\L$ in the above expression is the $n \rightarrow \infty$
limit of the expression (\ref{8a}),
\beq
\L\Phi = \left[ 2r {\hat e}_j (\Omega^{-1})^{ij}
{\del \Phi \over \del x^i}\right]_{n \rightarrow \infty}
\label{12c}
\eeq 

Given the fact that $\rho_0$ is a step function, see (\ref{step}), 
$\del \rho_0 /\del r^2$ is a $\delta$-function with support only at
the boundary of the droplet; as a result the $r$-integration
can be done and the action is entirely on the boundary of
the droplet. The $r$-integration will give an extra factor
of $n^{-k+1}$ for a compact $2k$ dimensional space. For the 
${\bf CP}^k$ case,
$N \sim n^k$, (\ref{dim}), so that the limit $n\rightarrow \infty$ gives a finite
prefactor to the action.
Also the potentials we choose will be such that
$\omega (r^2) \equiv {1 \over n} (\del V/\del r^2)$
is finite (not necessarily constant) as $n\rightarrow \infty$ \cite{KN2}. If the potential
does not have this property, then $\omega$ goes to zero
or infinity as $n\rightarrow \infty$. In the first case,
there are zero energy modes for the field $\Phi$,
indicating that the potential has not uniquely confined
the droplet to the shape initially chosen. In
the second case, no fluctuations of the droplet are energetically
allowed.  The final effective 
edge action is (up to a multiplicative constant) of 
the form
\beq
S_0 \approx -  \int_{\del\D} dt 
\left[ {\del \Phi \over \del t} (\L\Phi ) 
+\omega (\L\Phi )^2
\right] 
\label{12b}
\eeq
For ${\bf CP}^k$ the operator $\L$ in (\ref{12b}) can be written in terms of the complex coordinates as
\beq
\L = i (z \cdot {\del \over {\del z}}- \bz \cdot {\del \over {\del \bz}} )
\label{l}
\eeq

\section{Effective action with electromagnetic interactions}

In the presence of a weak electromagnetic field the effective action, as explained in section 2, is
\beq
S =S_0 + S_{A}
\label{s}
\eeq
where $S_0$ is the edge effective action given in (\ref{12a}) and $S_A$ is the $A$-dependent part of the action given by
\beqar
S_A & = & - N \int dt \Tr ~ \big( \hat{\rho}_0 \hat{ U}^{\dagger} \hat{ \A} \hat{U } \big) \nonumber \\
&=& -N \int dt \Tr ~\big( \hat{\rho}_0 \hat{\A} - i [\hat{\rho}_0,\hat{ \Phi} ] \hat{\A} +{1 \over 2} [\hat{\rho}_0,\hat{ \Phi} ] [\hat{\A}, \hat{\Phi}]  + \cdots \big) \nonumber \\
& = & -N\int dt d\mu ~ \big( \rho_0 * \A  + {1 \over n} \{ \rho_0, \Phi\} \A \big)~+ {\cal{O}}({1 \over n})  \nonumber \\
&=& - N \int dt d\mu ~ \big( \rho_0 * \A - {1 \over n}{{\del \rho_0} \over {\del r^2}}~\L \Phi ~\A \big) + {\cal{O}} ({1 \over n})
\label{sA}
\eeqar

The strategy then is to determine $\A$ as a function of $A_{\mu}$ so that the transformation (\ref{11a}) is a nonlinear realization of the usual gauge transformation for $A_{\mu}$, $\delta A_{\mu} = \del _{\mu} \Lambda$, where $\Lambda$ is the infinitesimal gauge parameter.

Using (\ref{com}) we find that the large $n$-limit of (\ref{11a}) is given by 
\beqar
\delta \A &  = & u^{\mu} \del_{\mu} \lambda + {1 \over n} (\Omega ^{-1})^{ij} \del_i \lambda \del _j \A \nonumber \\
u^{\mu} & = & (1, ~{1 \over n} (\Omega^{-1})^{ij} \del_jV)
\label{da}
\eeqar
The form of the above transformation suggests the following ansatz for $\A$ and $\lambda$ in terms of $A_{\mu}$ and $\Lambda$ :
\beqar
\A & = & a^{\mu} A_{\mu} + a^{\mu\nu} A_\mu A_\nu + a^{\mu\nu\lambda} A_{\mu}\del _{\nu} A_{\lambda} + \cdots \nonumber \\
\lambda & = &b \Lambda + b^{\mu\nu}  \del_{\mu} \Lambda A_{\nu} + \cdots 
\label{ansatz}
\eeqar
where $a$'s and $b$'s are to be determined. We compute $\delta\A$ using (\ref{ansatz}) and $\delta A_{\mu} = \del _{\mu} \Lambda$ and compare it to (\ref{da}) in order to determine these coefficients. We find the following results:
\beqar
&& b^{[\mu\nu]} \equiv {1 \over2} (b^{\mu\nu} - b^{\nu\mu})~,~~~~~~~~~~~b^{\{\mu\nu\}} = {1 \over 2} (b^{\mu\nu} + b^{\nu\mu}) \nonumber\\
&& a^0=b= {\rm constant} ~~~~~~~~~~~~~a^i  = b u^i + \kappa (\Omega ^{-1})^{ij}  \del_j \nonumber\\
&& b^{[0i]} =0~,~~~~~~b^{[ij]}= - {b^2 \over {2n}} (\Omega ^{-1})^{ij} \nonumber \\
&& a^{\{0\mu\}}={1 \over 2} u^{\lambda} \del_{\lambda}b^{\{0\mu\}}~,~~~~~~a^{\{ij\}} =  {1 \over 2} u^{\lambda} \del_{\lambda} b^{\{ij\}}+{b^2 \over {2n}} (\Omega ^{-1})^{ik} \del_k u^j  \nonumber \\
&& a^{000} = b^{00}~,~~~~~a^{i00} = a^{00i} = b^{\{0i\}}~,~~~a^{0i0} = b^{00} u^i \nonumber \\
&& a^{ij0} = b^{\{0i\}} u^j+{b^2 \over n} (\Omega^{-1})^{ij}~, ~~~a^{i0j}= b^{\{ij\}} - {b^2 \over {2n}} (\Omega^{-1})^{ij} ~,~~~a^{0ij} =u^i b^{\{0j\}} \nonumber \\
&& a^{ijk} = u^j b^{\{ik\}} - {b^2 \over {2n}} u^j (\Omega^{-1})^{ik} + { b^2 \over n} u^k (\Omega ^{-1})^{ij} 
\label{parameters}
\eeqar
where $b^{\{\mu\nu\}}$ and $\kappa$ are arbitrary. We then have
\beqar
 \A&  = &b u^\mu A_\mu - {b^2 \over {2n}} (\Omega^{-1})^{ij} \Big( A_i \del_0 A_j-2 A_i \del_j A_0 \Big)\nonumber \\
 & & -  {b^2 \over {2n}} (\Omega^{-1})^{ij} \Big( - \del_j u^k A_i A_k +  u^k A_i \del_k A_j - 2 u^k A_i \del_j A_k \Big)  \nonumber \\
& & -\kappa (\Omega^{-1})^{ij} F_{ij} + {1 \over 2} b^{\{\mu\nu\}} A_\mu u^{\lambda} \del_{\lambda} A_{\nu} + \cdots \\
\lambda  &= & b \Lambda - {b^2 \over {2n}} (\Omega^{-1})^{ij} \del_i \Lambda A_j + b^{\{\mu\nu\}} \del_\mu \Lambda A_\nu + \cdots 
\label{a}
\eeqar
We can set $b=1$ in the above expressions by absorbing $b$ in the definition of $A_\mu,~\Lambda$, namely $bA_\mu \rightarrow A_\mu,~b\Lambda \rightarrow \Lambda$. Further setting the undetermined parameters $\kappa, b_{\{\mu\nu\}}$ to zero, we obtain the ``minimal" solution which is given by
\beqar
 \A & = & u^\mu A_\mu - {1 \over {2n}} (\Omega^{-1})^{ij} \Big( A_i \del_0 A_j-2 A_i \del_j A_0 \Big) \nonumber \\
& & -   {1 \over {2n}} (\Omega^{-1})^{ij} \Big( - \del_j u^k A_i A_k +  u^k A_i \del_k A_j - 2 u^k A_i \del_j A_k \Big)   \\
\lambda & = & \Lambda - {1 \over {2n}} (\Omega^{-1})^{ij} \del_i \Lambda A_j 
\label{minimal}
\eeqar
In the case of the two-dimensional plane, where  ${1 \over n} (\Omega^{-1})^{ij} \rightarrow {1 \over B} \epsilon^{0ij}$, (38), (39) agree with the expressions in \cite{sakita2}.

Before proceeding with the calculation of $S_A$ in (\ref{sA}) it is interesting to point out the relation of (38)-(\ref{minimal}) to the Seiberg-Witten map \cite{seib1, seib2}. In that context $\A$ is the time component of the noncommutative gauge field and in the absence of the confining potential $V$, when $u^i=0$, the transformations (40) is exactly the one defining the Seiberg-Witten map between the noncommutative gauge field $\A$ and the commutative gauge field $A_\mu$, where our $\Omega^{-1}$ is identified in terms of the noncommutative parameters $\theta^{ij}$ as 
\beq
[x_i,~x_j]_* \equiv i \theta^{ij} =   - {i \over n} (\Omega^{-1})^{ij}
\label{id}
\eeq
Our expression (40) for $V=0$ is essentially the Seiberg-Witten mapping for $\A$ upto linear order in $\Omega ^{-1}$, which is sufficient for large $n$. One can easily show that our expression in the presence of the confining potential, can be derived by considering the Seiberg-Witten map to second order in $\Omega ^{-1}$ \cite{seib2}, by replacing $\A \rightarrow \A + V,~ A_0 \rightarrow A_0 + V$ and by keeping up to $1/n$-order terms.

\subsection{ Calculation of $S_A$}

Using (\ref{star}), (\ref{omega}) we find that the large $n$-limit of $S_A$ is given by 
\beq
S_A = -N \int dt d\mu ~ \Big[ \rho_0 \A + {1 \over {2n}} {{\del \rho_0} \over {\del r^2}} (1 + \bz \cdot z)^2 \Big(z \cdot {\del \over {\del z}} + \bz \cdot {\del \over {\del \bz}} \Big) ~u^\mu A_\mu   -{ 1\over n} {{\del \rho_0} \over {\del r^2}} \L \Phi ~u^\mu A_\mu \Big]
\label{sa}
\eeq
where $\A$ can be also written as
\beqar
\A  = && u^\mu A_\mu - {1 \over {2n}} (\Omega^{-1})^{ij} \Big( A_i \del_0 A_j-2 A_i \del_j A_0-\del_i(A_0 A_j)\Big) \nonumber \\
&&- {1 \over {2n}} (\Omega^{-1})^{ij} \del_i \Big( u^\mu A_\mu A_j \Big) \nonumber\\
&&- {1 \over {2n}}  (\Omega^{-1})^{ij} u^k \Big(  A_i \del_k A_j - A_i \del_j A_k + \del_j A_i A_k \Big)
\label{a1}
\eeqar
The above expression for $\A$ is identical to (40), but it is more convenient in order to display the bulk and boundary contributions of $S_A$. In particular the term $(\Omega^{-1})^{ij} \del _i (A_0 A_j)$ has been added and subtracted in order that the bulk action produces the full Chern-Simons 3-form $AdA \sim A_{[ \mu} \del _\nu A_{\lambda ]}$, as we shall see in the following. The bulk contribution for $S_A$ is
\beqar
S_{A,{\rm bulk}} = -N \int dt d\mu ~\rho_0 &&  \Big[  u^\mu A_\mu - {1 \over {2n}} (\Omega^{-1})^{ij} \Big(A_i \del_0 A_j-2 A_i \del_j A_0-\del_i(A_0 A_j)\Big) \nonumber \\
&& -  {1 \over {2n}}  (\Omega^{-1})^{ij} u^k \Big( A_i \del_k A_j - A_i \del_j A_k + \del_j A_i A_k \Big) \Big]
\label{bulk}
\eeqar
Using the fact that for ${\bf CP}^k$
\beq
d\mu = {k! \over {\pi ^k}} {{d^{2k} x} \over {(1 + \bz \cdot z)^{k+1}} }= {{d^{2k} x} \over {(4 \pi)^k}} {\rm det} \Omega 
\label{measure}
\eeq
we find 
\beq
d\mu~(\Omega^{-1})^{ij}= -{{2k} \over {(4 \pi)^k}} \epsilon^{iji_1j_1 \cdots i_{k-1}j_{k-1}} \Omega_{i_1j_1} \cdots \Omega_{i_{k-1}j_{k-1}} ~d^{2k} x 
\label{dmuo}
\eeq
Using (\ref{dmuo}) we find that the second term produces a Chern-Simons type term of the form
\beq
{{Nk} \over n} \int {{d^{2k} x} \over {(4 \pi)^k}} ~\rho_0 ~\epsilon^{\mu\nu\lambda \alpha_1\beta_1\cdots \alpha_{k-1}\beta_{k-1}} A_{\mu}\del_{\nu} A_{\lambda} \Omega_{\alpha_{1}\beta_{1}} \cdots \Omega_{a_{k-1}\beta_{k-1}}
\label{secondb}
\eeq
In the case of ${\bf CP}^1$, $k=1$, this is the usual Chern-Simons term. For $\bf{CP}^k$, $k>1$, this is the so-called K\"ahler-Chern-Simons term \cite{nair}.

The third term in (\ref{bulk}) produces another bulk term of the form

\beq
- {{Nk} \over n} \int {{d^{2k} x} \over {(4 \pi)^k}} \rho_0  \epsilon^{ij i_1 j_1 \cdots i_{k-1} j_{k-1}} u^k \big[ A_i F_{kj} - {1 \over 2} A_k F_{ij} \big]  \Omega_{i_1j_1} \cdots \Omega_{i_{k-1} j_{k-1}} 
\label{thirdb}
\eeq
After some algebra one can show that this can be written as 
\beqar
& & {{Nk} \over n} (k-1) \int {{d^{2k} x} \over {(4 \pi)^k}} \rho_0  \epsilon^{ikj j_1 \cdots i_{k-1} j_{k-1}} 
u^l A_i F_{kj} \Omega_{l j_1} \cdots \Omega_{i_{k-1} j_{k-1}} \nonumber \\
&& = - {{Nk} \over n} (k-1) \int {{d^{2k} x} \over {(4 \pi)^k}} \rho_0  \epsilon^{ikjli_2j_2 \cdots i_{k-1} j_{k-1}} 
A_i F_{kj} { 1 \over n} \del_l V \Omega_{i_2 j_2} \cdots \Omega_{i_{k-1} j_{k-1}} 
\label{3b}
\eeqar
For ${\bf CP}^1$ this term is obviously zero and the only bulk term besides the first term in (\ref{bulk}) is the Chern-Simons term as expected. However, for ${\bf CP}^k$, $k>1$, (\ref{3b}) produces an additional bulk term.

The edge contribution of $S_A$ is
\beqar
S_{A,{\rm edge}} = && {N \over {2n}} \int dt  d\mu  \rho_0 (\Omega^{-1})^{ij} \del_i (u^\mu A_\mu A_j) \nonumber \\
&- & {N \over {2n}} \int dt d\mu {{\del \rho_0} \over {\del r^2}} (1 + \bz \cdot z)^2 \Big(z \cdot {\del \over {\del z}} + \bz \cdot {\del \over {\del \bz}} \Big) ~u^\mu A_\mu \nonumber \\
& + &{ N \over n} \int dt d\mu {{\del \rho_0} \over {\del r^2}} \L \Phi ~ u^\mu A_\mu 
\label{edge}
\eeqar
Given the fact that $\del_i \Big[ (\Omega^{-1})^{ij} \det \Omega \Big] =0$ and $\rho_0$ corresponds to a spherical droplet $\rho_0 = \rho_0 (r^2)$, one can easily see that the first term in (\ref{edge}) is a boundary term.

Putting everything together we get
\beq
S_A = S_{A, {\rm bulk}} +  S_{A, {\rm edge}}
\label{sat}
\eeq
\beqar
S_{A,{\rm bulk}} =&-& N \int dt d\mu \rho_0 u^\mu A_\mu \nonumber\\
&+& {{Nk} \over n} \int {{d^{2k} x} \over {(4 \pi)^k}} \rho_0 \epsilon^{\mu\nu\lambda \alpha_1\beta_1\cdots \alpha_{k-1}\beta_{k-1}} A_{\mu}\del_{\nu} A_{\lambda} \Omega_{\alpha_{1}\beta_{1}} \cdots \Omega_{a_{k-1}\beta_{k-1}} \nonumber \\
&-& {{Nk} \over n} (k-1) \int {{d^{2k} x} \over {(4 \pi)^k}} \rho_0  \epsilon^{ikjli_2j_2 \cdots i_{k-1} j_{k-1}} 
A_i F_{kj} { 1 \over n} \del _l V \Omega_{i_2 j_2} \cdots \Omega_{i_{k-1} j_{k-1}} 
\label{bulkt} \\
S_{A,{\rm edge}}=&-& {N \over {2n}} \int dt  d\mu  \rho_0 (\Omega^{-1})^{ij} \del_i (u^\mu A_\mu A_j)\nonumber \\
&-&{N \over {2n}} \int dt d\mu {{\del \rho_0} \over {\del r^2}} (1+\bz \cdot z)^2 \Big(z\cdot \del + \bz \cdot \bar{\del}\Big) u^\mu A_\mu \nonumber \\
& + & { N \over n} \int dt d\mu  {{\del \rho_0} \over {\del r^2}} \L \Phi ~ u^\mu A_\mu 
\label{edget}
\eeqar
The $V$-dependent terms in $S_A$ can be thought of as arising from the fact that the derivative of the confining potential acts as an effective electric field. 

Using (\ref{bulkt}) we can derive the Hall current as
\beqar
J^0 & = & { {\delta S_{A, {\rm bulk}}} \over {\delta A_0}} = -{N \over {(4 \pi)^k}} \rho_0 \Big[ \det {\Omega} - {k \over n} \epsilon^{iji_1j_1\cdots i_{k-1}j_{k-1}} F_{ij} \Omega_{i_1j_1}\cdots\Omega_{i_{k-1}j_{k-1}} \Big]  \label{j0}\\
J^i & = &  {{\delta S_{A, {\rm bulk}}} \over {\delta A_i}} = -{N \over {(4 \pi)^k}} \rho_0 \Big[ \det \Omega ~ u^i +{ 2k \over n} \epsilon ^{iji_1j_1\cdots i_{k-1}j_{k-1}} F_{0j}  {\Omega}_{i_1j_1} \cdots {\Omega} _{i_{k-1}j_{k-1}} \nonumber \\
&&+ {{2k(k-1)} \over n} \epsilon ^{ijkli_2j_2\cdots i_{k-1}j_{k-1}} {1 \over n} \del _j V F_{kl} ~ {\Omega}_{i_2j_2} \cdots {\Omega} _{i_{k-1}j_{k-1}} \Big]
\label{ji}
\eeqar
The above expressions for the Hall current can be compactly written, up to higher order terms in $1/n$, as
\beqar
J^0 & =& - {N \over {(4 \pi)^k}} \rho_0 \det \tilde {\Omega}  \nonumber \\
J^i & = & -{N \over {(4 \pi)^k}} \rho_0 { 2k \over n} \epsilon ^{iji_1j_1\cdots i_{k-1}j_{k-1}} \tilde {F}_{0j} \tilde {\Omega}_{i_1j_1} \cdots \tilde{\Omega} _{i_{k-1}j_{k-1}} 
\label{current}
\eeqar
where
\beqar
\tilde{A_i} & = & A_i ,~~~~~~~~~~~~~~\tilde{A_0}= A_0 + V \nonumber \\
\tilde{\Omega}_{ij} & = & \Omega _{ij} -{1 \over n} F_{ij} 
\label{tilde}
\eeqar
In fact, using (\ref{tilde}) one can show that $S_{A,{\rm bulk}}$ can be written, up to a boundary term,  as a K\"ahler-Chern-Simons action in terms of $\tilde{A},~\tilde{\Omega}$,
\beqar
S_{A, {\rm bulk}} = &-& N \int dt d\mu ~ \rho_0 A_0 \nonumber \\
&+& {{Nk} \over n} \int {{d^{2k} x} \over {(4 \pi)^k}} ~ \rho_0 ~ \epsilon^{\mu\nu\lambda \alpha_1\beta_1\cdots \alpha_{k-1}\beta_{k-1}} \tilde{A}_{\mu}\del_{\nu} \tilde{A}_{\lambda} \tilde{\Omega}_{\alpha_{1}\beta_{1}} \cdots \tilde{\Omega}_{a_{k-1}\beta_{k-1}} 
\label{cs}
\eeqar

\section{Gauge invariance of effective action}

One of the important points to be emphasized in our approach of constructing the effective action for a Hall droplet in the presence of a weak electromagnetic field is that it incorporates gauge invariance right from the beginning, eqs. (\ref{10})-(\ref{12}).

In this section we shall discuss the gauge transformation of the edge and bulk parts of the action and explicitly demonstrate its gauge invariance. As we showed earlier the total effective action splits in three parts
\beq
S= S_0 + S_{A,{\rm edge}} + S_{A,{\rm bulk}}
\eeq
where $S_0$ describes the edge dynamics of the droplet in the absence of electromagnetic interactions, and $S_{A,{\rm edge}}$ and  $S_{A,{\rm bulk}}$ are the edge and bulk contributions of the action which depend explicitly on the electromagnetic field $A_\mu$.
\vskip .2in
\noindent
$\underline{{\rm Gauge~ transformation ~of ~}S_0}$
\vskip .1in

The gauge transformation of the matter field is given in (\ref{11}). One can show that this implies the following gauge transformation for $\Phi$:
\beq
\delta \Phi = -\lambda +{1 \over {2n}} \{ \lambda,~\Phi\} + \cdots
\label{deltaphi}
\eeq
Using (\ref{minimal}), we find that to lowest order in $1/n$ we have \footnote{Setting $b=1$ in deriving (\ref{minimal}) is equivalent to fixing the charge for the field $\Phi$.}
\beq
\delta \Phi = -\Lambda + \O ({1 \over n})
\label{dp}
\eeq
Using (\ref{dp}) we find that the gauge transformation of $S_0$ is 
\beq
\delta S_0 = {N \over {n}} \int dt d\mu {{\del \rho_0} \over {\del r^2}} \left[ - \del_t \Lambda \L \Phi  - {1 \over n} {{\del V} \over {\del r^2}} \L \Lambda ~ \L \Phi \right]
\label{ds0}
\eeq
\vskip .2in
\noindent
$\underline{{\rm Gauge~transformation~of~}S_{A,{\rm bulk}}}$
\vskip .1in

The gauge transformation of the combination $u^\mu A_\mu$ is
\beq
\delta ( u^\mu A_\mu) = \del _t {\Lambda} + u^i \del_i \Lambda = \del _t {\Lambda} + {1 \over n} {{\del V} \over {\del r^2}}  \L \Lambda
\label{duA}
\eeq
Based on this one can show  that the first term in the expression (\ref{bulkt}) for $S_{A,{\rm bulk}}$ is gauge invariant. The third term is also gauge invariant. The gauge non-invariance of $S_{A,{\rm bulk}}$ is due to the second term, the K\"ahler-Chern-Simons term. In particular we find
\beq
\delta S_{A,{\rm bulk}} = - {N \over {2n}} \int dt d\mu  \rho_0 ~ (\Omega^{-1})^{ij} \del_i (\del _t \Lambda A_j)~+~ {N \over {2n}} \int dt d\mu {{\del \rho_0} \over {\del r^2}} ~ \L \Lambda A_0 
\label{dsab}
\eeq

\noindent
$\underline{{\rm Gauge~transformation~of~}S_{A,{\rm edge}}}$
\vskip .1in

Using (\ref{duA}) one can show that the second term in the expression (\ref{edget}) for $S_{A,{\rm edge}}$ is gauge invariant. The gauge non-invariance of $S_{A,{\rm edge}}$ results from the gauge transformation of the first and third term. In particular we find,
\beq
\delta S_{A,{\rm edge}} ={N \over {2n}} \int dt d\mu  \rho_0 ~ (\Omega^{-1})^{ij} \del_i (\del _t \Lambda A_j)~+~{N \over {2n}} \int dt d\mu {{\del \rho_0} \over {\del r^2}}~  \Big[ - \L \Lambda  A_0 + 2 \L \Phi  \del _t \Lambda + {2 \over n} {{\del V} \over {\del r^2}} ~\L \Phi ~ \L \Lambda \Big] 
\label{dsae}
\eeq
\vskip .2in
In deriving (\ref{ds0}), (\ref{dsab}), (\ref{dsae}) we used the fact that $\rho_0$ is time independent, and integrals of total time derivative and total angular derivative terms are zero. 

Using (\ref{dsae}) and (\ref{dsab}) we find that the gauge transformation of the total $S_A$ is 
\beq
\delta S_A = {N \over n} \int dt d\mu {{\del \rho_0} \over {\del r^2}} \big[ \L \Phi~\del _t  \Lambda + {1 \over n} {{\del V} \over {\del r^2}} \L \Phi ~ \L \Lambda \big]
\label{dsA}
\eeq
Adding (\ref{ds0}) and (\ref{dsA}) is is easy to see that the total effective action is gauge invariant.
\beq
\delta S = \delta S_0 + \delta S_{A,{\rm edge}} + \delta S_{A,{\rm bulk}} =0
\label{ds}
\eeq

\section{Conclusions and comments}

Using the fact that electromagnetic interactions of spinless fermions in the lowest Landau level can be described in terms of a $W_N$-gauge theory, where the $W_N$ transformations are nonlinear realizations of $U(1)$ gauge transformations, we were able to construct the effective action describing electromagnetic interactions of a higher dimensional quantum Hall droplet. This method has the advantage of deriving simultaneously the bulk and the boundary parts of the action, and demonstrating the explicit gauge invariance of the total action. The bulk term is a Chern-Simons type action whose anomaly is cancelled by the boundary term.

Our results for the bulk action (\ref{cs}) and the Hall current (\ref{current}) can be used to derive an effective topological field theory for higher dimensional Hall droplets as in \cite{toumbas}. In the case of ${\bf CP}^k$ with $U(1)$ background magnetic field we obtain a K\"ahler-Chern-Simons action instead of the usual $U(1)$ Chern-Simons action in (2k+1) dimensions.

Since in the absence of electromagnetic interactions the dynamics of Hall droplet is given by an Abelian bosonic chiral action, this method can also be thought of as providing, beyond the quantum Hall physics, a consistent way of gauging a bosonic chiral action in higher dimensions. The choice of the "minimal" solution, eqs. (40-41), corresponds to a minimal gauge coupling for the chiral action.

The extension of these results to the case of higher dimensional, non-Abelian quantum Hall droplets \cite{KN3} and non-Abelian external fields is interesting and is currently under consideration.

\vskip .3in
\noindent
{\bf Acknowledgements}

We acknowledge useful discussions with V.P. Nair. This work was supported in part by the National Science Foundation under grant number PHY-0140262 and by a PSC-CUNY grant.

\end{document}